\documentclass[11pt]{article}

\usepackage[dvipsnames]{xcolor}
\usepackage{color} 
\usepackage{amsmath}
\usepackage{amsfonts}
\usepackage{amssymb,amsmath,amsthm}
\usepackage{graphicx}
\usepackage{subfigure}
\usepackage{longtable}
\usepackage{multirow}
\usepackage{url}
\usepackage{ulem}

\usepackage[colorlinks=true,citecolor=blue,urlcolor=blue,linkcolor=blue]{hyperref}
\usepackage{caption}
\usepackage{lipsum}

\newcommand\dd{\hbox{d}}
\newcommand{\E}{\mathbb{E}}
\newcommand{\R}{\mathbb{R}}
\newcommand{\cR}{\mathcal{R}}
\newcommand{\im}{{\mathrm{i}}}
\newcommand{\bX}{{\boldsymbol{X}}}
\newcommand{\bY}{{\boldsymbol{Y}}}
\newcommand{\V}{\mathcal{V}}
\newcommand{\Cov}{\hbox{Cov}}
\newcommand{\Var}{\hbox{Var}}

\def\tinyone{\fontsize{1pt}{1pt}\selectfont}

\begin{document}

\title{\Large\bf Distance Correlation: A New Tool for Detecting Association \\ and Measuring Correlation Between Data Sets}

\author{{\large\bf{Donald St. P. Richards}}\footnote{
Donald Richards is a professor of Statistics at Penn State University.  His e-mail address is: richards@stat.psu.edu  
\hfill\break
{\tinyone{.}}\quad This research was supported in part by National Science Foundation grants AST-0908440 and DMS-1309808, and by a Romberg Guest Professorship at the University of Heidelberg Graduate School for Mathematical and Computational Methods in the Sciences, funded by German Universities Excellence Initiative grant GSC 220/2.
}}

\date{January 1, 2017}

\maketitle

The difficulties of detecting association, measuring correlation, and establishing cause and effect have fascinated mankind since time immemorial.  Democritus, the Greek philosopher, underscored well the importance and the difficulty of proving causality when he wrote, ``I would rather discover one cause than gain the kingdom of Persia'' \cite[p. 104]{Freeman}.  

To address the difficulties of relating cause and effect, statisticians have developed many inferential techniques.  Perhaps the most well-known method stems from Karl Pearson's coefficient of correlation, which Pearson introduced in the late 19th century based on ideas of Francis Galton.

Let $X$ and $Y$ be non-trivial scalar random variables.  Let $\E(X)$ the {\it mean} of $X$, $\Var(X) = \E(X^2) - [\E(X)]^2$ be the {\it variance} of $X$, and $\Cov(X,Y) = \E(XY) - \E(X)\E(Y)$ be the {\it covariance} between $X$ and $Y$.  
%If $X$ and $Y$ are independent then $\E(XY) = \E(X)\E(Y)$ and therefore $\Cov(X,Y) = 0$.  
The {\it Pearson correlation coefficient} between $X$ and $Y$ is defined to be 
$$
\frac{\Cov(X,Y)}{\sqrt{\Var(X)} \cdot \sqrt{\Var(Y)}}.
$$
If $X$ and $Y$ are mutually independent then $\Cov(X,Y) = 0$; however, the converse does not hold.  If there is a plausibly close-to-linear relationship between $X$ and $Y$ then the Pearson coefficient measures the strength of that linear relationship.  

On drawing from the joint distribution of $(X,Y)$ a random sample $(x_1,y_1),\ldots,(x_N,y_N)$ the {\it empirical} Pearson correlation coefficient is 
\begin{equation*}
%\label{eq:empirical-Pearson}
\frac{\sum_{i=1}^N (x_i-\bar{x})(y_i-\bar{y})}{\sqrt{\sum_{i=1}^N (x_i-\bar{x})^2} \cdot \sqrt{\sum_{i=1}^N (y_i-\bar{y})^2}} \, ,
\end{equation*}
where $\bar{x} = N^{-1}\sum_{i=1}^N x_i$ and $\bar{y} = N^{-1}\sum_{i=1}^N y_i$ are the respective sample means.  The classical Pearson correlation has the attractive features of invariance under shifting or dilations of $X$ or $Y$, and it also is well-known for its amenability to graphical methods of data analysis.  

The Pearson coefficient applies only to scalar random variables, however, and it is inapplicable generally if an existing relationship between $X$ and $Y$ is highly non-linear; this has led to the amusing enumeration of correlations between many pairs of unrelated variables, e.g., ``Median salaries of college faculty'' and ``Annual liquor sales in college towns.''  

Let $p$ and $q$ be positive integers, and let $X = (X_1,\ldots,X_p) \in \R^p$ and $Y = (Y_1,\ldots,Y_q) \in \R^q$ be random vectors.  For $s = (s_1,\ldots,s_p) \in \R^p$, the norm $\|s\| = (s_1^2+\cdots+s_p^2)^{1/2}$ denotes the standard Euclidean norm on $\R^p$, and $\langle s,X\rangle = s_1X_1+\cdots+s_pX_p$ is the standard inner product between $s$ and $X$.  
Similarly, for $t \in \R^q$, we denote by $\|t\|$ and $\langle t,Y\rangle$ the corresponding Euclidean norm and inner product on $\R^q$.  

Define for $(s,t) \in \R^p \times \R^q$ the {\it joint characteristic function} of $(X,Y)$, 
$$
\phi_{X,Y}(s,t) = \E \exp\left[\im\langle s,X\rangle + \im\langle t,Y\rangle\right],
$$
$\im = \sqrt{-1}$, and the {\it marginal characteristic functions} of $X$ and $Y$, 
$
\phi_X(s) = \E \exp\left[\im \langle s,X\rangle\right]
$ 
and 
$
\phi_Y(t) = \E \exp\left[\im \langle t,Y\rangle\right],
$ 
respectively.  It is well-known that $X$ and $Y$ are mutually independent if and only if $\phi_{X,Y}(s,t) = \phi_X(s)\phi_Y(t)$ for all $(s,t) \in \R^p \times \R^q$.  
Sz\'ekely, {\it et al.} \cite{szekely07,szekely09} (see also Feuerverger \cite{Feuerveger}) defined, for random vectors $X$ and $Y$ with finite first moment, the {\it distance covariance}, 
\begin{equation}
\label{dcov}
\V(X,Y) =  \left(\frac{1}{c_{p}c_{q}} 
\int \frac{|\phi_{X,Y}(s,t)-\phi_{X}(s)\phi_{Y}(t)|^{2}}{\|s\|^{p+1} \, \|t\|^{q+1}} \, \dd s \, \dd t\right)^{1/2},
\end{equation}
where the integral is with respect to Lebesgue measure on ${\R^p \times \R^q}$ and 
\begin{equation}
\label{eq:cp}
c_p = \frac{\pi^{(p+1)/2}}{\Gamma\big((p+1)/2\big)}.
\end{equation}
The {\it distance correlation coefficient} between $X$ and $Y$ is defined as 
\begin{equation*}
%\label{dcorr}
\cR(X,Y) = \frac{\V(X,Y)}{\sqrt{\V(X,X)} \cdot \sqrt{\V(Y,Y)}}
\end{equation*}
if $\V(X,X), \V(Y,Y) > 0$; otherwise, $\cR(X,Y)$ is defined to be $0$.  Thus, the distance correlation is defined for vectors $X$ and $Y$ of arbitrary dimension.  Moreover, it follows from (\ref{dcov}) that $X$ and $Y$ are mutually independent if and only if $\cR(X,Y) = 0$.  These properties provide advantages of $\cR(X,Y)$ over the Pearson coefficient and other classical measures of correlation.  

For a given pair of jointly distributed random vectors $(X,Y)$, it is a non-trivial problem to calculate $\V(X,Y)$.  We shall describe the recent results of Dueck, {\it et al.} \cite{dueck16}, in which $\V(X,Y)$ have been calculated for the class of Lancaster probability distributions.  

For a random sample $(X_1,Y_1),\ldots,(X_N,Y_N)$ from the joint distribution of $(X,Y)$, set $\bX = [X_1,\ldots,X_N]$ and $\bY = [Y_1,\ldots,Y_N]$ and define the {\it joint empirical characteristic function}, 
$$
\phi_{\bX,\bY}^N(s,t) = \frac1N \sum_{j=1}^N 
\exp\big[ \im \langle s,X_j \rangle_p + \im \langle t,Y_j \rangle \big],
$$
$(s,t) \in \R^p \times \R^q$.  Writing $\phi^N_\bX(s) = \phi_{\bX,\bY}^N(s,0)$ and $\phi^N_\bY(t) = \phi_{\bX,\bY}^N(0,t)$ for the corresponding marginal empirical characteristic functions, we define the {\it empirical distance covariance} by
\begin{equation}
\label{eq:empiricaldcov}
\V_N(\bX,\bY) = \left(\frac{1}{c_p c_q} \int 
\frac{|\phi_{\bX,\bY}^N(s,t)-\phi_\bX^N(s)\phi_\bY^N(t)|^2}{\|s\|^{p+1} \, 
\|t\|^{q+1}} \dd s \, \dd t\right)^{1/2},  
\end{equation}
where $c_p$ is given in (\ref{eq:cp}).  Sz\'ekely, {\it et al.} \cite{szekely07,szekely09} (and also earlier, Feuerverger \cite{Feuerveger}), in a {\it tour de force}, proved that 
\begin{equation}
\label{eq:sample.dcov}
[\V_N(\bX,\bY)]^2 = \frac{1}{N^2} \sum_{k,l=1}^N A_{kl} B_{kl}
\end{equation}
where, for $1 \le k,l \le N$, $a_{kl} = \|X_k-X_l\|_p$, 
$$
\bar{a}_{k\boldsymbol\cdot} = \frac{1}{N} \sum_{l=1}^N a_{kl}, 
\ \ 
\bar{a}_{\boldsymbol\cdot l} = \frac{1}{N} \sum_{k=1}^N a_{kl}, 
\ \
\bar{a}_{\boldsymbol\cdot\boldsymbol\cdot} = 
\frac{1}{N^2} \sum_{k,l=1}^N a_{kl},
$$ 
$$
A_{kl} = a_{kl} - \bar{a}_{k\boldsymbol\cdot} - \bar{a}_{\boldsymbol\cdot l} 
         + \bar{a}_{\boldsymbol\cdot\boldsymbol\cdot},
$$
and similarly for $b_{kl} = \|Y_k-Y_l\|$, $\bar{b}_{k\boldsymbol\cdot}$, $\bar{b}_{\boldsymbol\cdot l}$, and $B_{kl}$.  Hence, the empirical quantity, $\V_N(\bX,\bY)$, although defined by the formidable integral (\ref{eq:empiricaldcov}), can be calculated directly using (\ref{eq:sample.dcov}).  

The {\it empirical distance correlation} for the observed data $(\bX,\bY)$ is defined as 
\begin{equation*}
%\label{eq:empirical-dcor}
\cR_N(\bX,\bY) = \frac{\V_N(\bX,\bY)}{\sqrt{\V_N(\bX)} \cdot \sqrt{\V_N(\bY)}}
\end{equation*}
if $\V_N(\bX), \V_N(\bY) > 0$; otherwise, $\cR_N(\bX,\bY)$ is defined to be $0$.  

The distance correlation coefficient has now been applied in many contexts and it has been found to exhibit higher {\it statistical power} (i.e., fewer false positives) than the Pearson coefficient, to find nonlinear associations that were undetected by the Pearson coefficient, and to locate smaller sets of variables that provide equivalent statistical information.  

In the field of astrophysics, large amounts of data are collected and stored in publicly-available repositories.  The COMBO-17 database \cite{wol03a}, for instance, provides numerical measurements on many astrophysical variables for more than 63000 galaxies, stars, quasars, and unclassified objects in the Chandra Deep Field South region of the Sky, with brightness measurements over a wide range of redshifts.  Current understanding of galaxy formation and evolution is sensitive to the relationships between astrophysical variables, so it is essential in astrophysics to be able to detect and verify associations between variables.  

\begin{figure*}[t]
\captionsetup{width=0.85\textwidth}
\center
\includegraphics[scale=0.29]{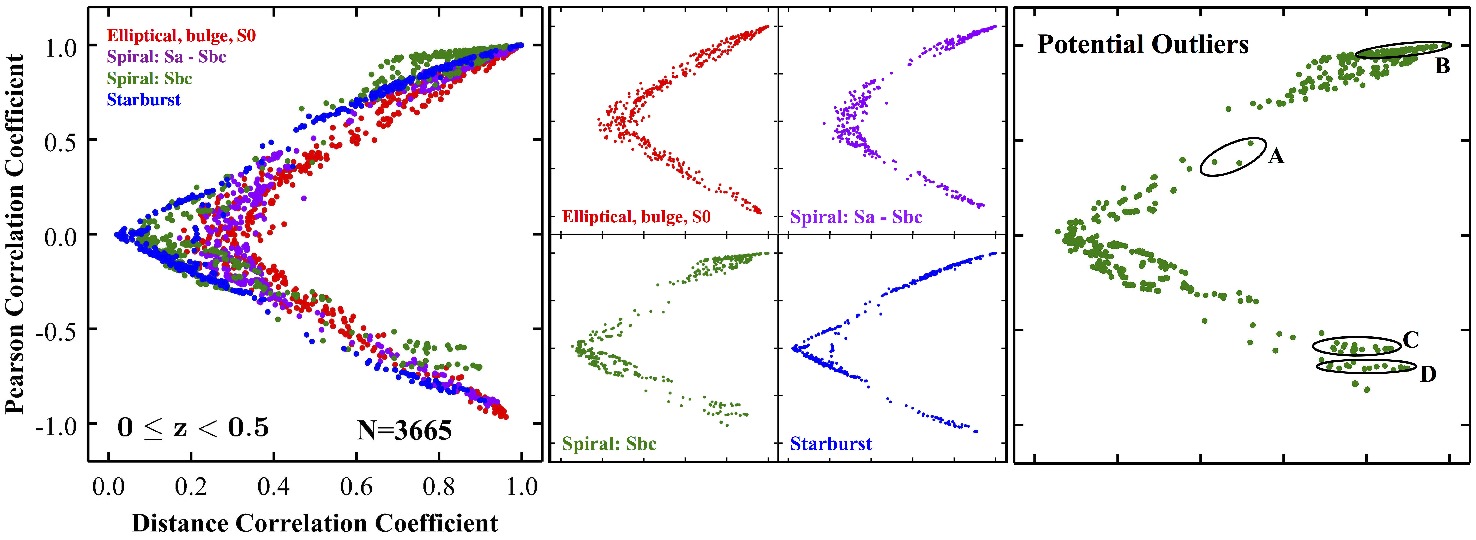}
\caption{Pearson correlation coefficient {\it vs.} distance correlation coefficient for 528 pairs of variables, calculated for galaxies with redshift $0 \leq z < 0.5$.  The subplots for each galaxy type are shown in four middle frames together with the superposition of the four subplots in the large left frame.   The locations of potential outlier pairs in the scatter plot for {\bf\color{OliveGreen}Type 3 (Spiral: Sbc)} galaxies is also shown in the large right frame.
}
\label{fig1}
\end{figure*}

Mercedes Richards, {\it et al.} \cite{martinez14,richards14} applied the distance correlation method to 33 variables measured on 15352 galaxies in the COMBO-17 database.  For each of the $\binom{33}{2} = 528$ pairs of variables, the Pearson and distance correlation coefficients were computed and graphed in Figure \ref{fig1} for the subset of galaxies with redshift $z \in [0,0.5)$.  It was determined that, for given values of the Pearson coefficient, the distance correlation exhibited a greater ability than other measures of correlation to resolve astrophysical data into highly concentrated horseshoe- or V-shapes. These results were observed over a range of redshifts beyond the Local Universe and for galaxies ranging in type from elliptical to spiral.  

The greater ability of the distance correlation to resolve data into well-defined horseshoe- or V-shapes led in turn to more accurate classification of galaxies and identification of outlying pairs of variables.  As seen in Figure \ref{fig1}, the {\color{Fuchsia} \bf Type 2} and {\color{OliveGreen} \bf Type 3} groups of spiral galaxies are likely to be contaminated substantially by {\color{blue} \bf Type 4} starburst galaxies, confirming earlier findings of other astrophysicists.  Our study found further evidence of this contamination from the distribution of points in the V-shaped scatter plots for galaxies of Types {\color{Fuchsia} \bf 2} or {\bf\color{OliveGreen} \bf 3}.

%Figure 1

I will also discuss data arising in the national discussion of the relationship between homicide rates and the strength of state gun laws.\footnote{\color{red}As sure as my first name is ``Donald'', this part of the talk will be {\it Huge}!}  A {\sl Washington Post} columnist has claimed that there is ``Zero correlation between state homicide rate and state gun laws'' \cite{volokh}.  Indeed, calculation of the corresponding empirical Pearson coefficient detects no statistically significant relationship between those variables.  However, a distance correlation analysis of the data discovers overwhelming evidence that, when the states are partitioned by region or by median population density, there is a strong relationship between the two variables.

There is also the intriguing formula (\ref{eq:sample.dcov}) for the empirical distance covariance.  Behind that formula lies a remarkable singular integral: For $x \in \R^p$ and $\hbox{Re}(\alpha) \in (0,2)$, 
$$
\int_{\R^p} \frac{1 - e^{\im \langle s,x\rangle}}{\|s\|^{p+\alpha}} \dd s = \frac{2 \pi^{p/2} \Gamma\big(1-\tfrac12 \alpha\big)}{\alpha 2^\alpha \Gamma\big(\tfrac12(p+\alpha)\big)} \|x\|^\alpha,
$$
with absolute convergence for all $x$.  We shall describe generalizations of this singular integral arising from truncated Maclaurin expansions of the cosine function and in the theory of spherical functions on symmetric cones.

{\small

\bibliographystyle{ims}

}

\end{document}